\documentclass[prl,twocolumn,aps,a4paper,superscriptaddress,showpacs]{revtex4}
\usepackage{amsmath}
\usepackage[dvips]{graphicx}
\begin{document}

\newcommand\be{\begin{equation}}
\newcommand\ee{\end{equation}}
\newcommand\bea{\begin{eqnarray}}
\newcommand\eea{\end{eqnarray}}
\newcommand\bseq{\begin{subequations}} %solo con amsmath
\newcommand\eseq{\end{subequations}}
\newcommand\bcas{\begin{cases}}
\newcommand\ecas{\end{cases}}
\newcommand{\p}{\partial}
\newcommand{\f}{\frac}

\title{The Big-Bang Singularity\\ in the framework of a Generalized Uncertainty Principle}

\author{Marco Valerio Battisti}
\email{battisti@icra.it}
\affiliation{ICRA - International Center for Relativistic Astrophysics}
\affiliation{Dipartimento di Fisica (G9), Universit\`a di Roma ``La Sapienza'' P.le A. Moro 5, 00185 Rome, Italy}
\author{Giovanni Montani}
\email{montani@icra.it} 
\affiliation{ICRA - International Center for Relativistic Astrophysics}
\affiliation{Dipartimento di Fisica (G9), Universit\`a di Roma ``La Sapienza'' P.le A. Moro 5, 00185 Rome, Italy}
\affiliation{ENEA C.R. Frascati (Dipartimento F.P.N.), Via Enrico Fermi 45, 00044 Frascati, Rome, Italy}

%\today

\begin{abstract}
We analyze the quantum dynamics of the Friedmann-Robertson-Walker Universe in the context of a Generalized Uncertainty Principle. Since the isotropic Universe dynamics resembles that of a one-dimensional particle, we quantize it with the commutation relations associated to an extended formulation of the Heisenberg algebra. The evolution of the system is described in terms of a massless scalar field taken as a relational time. We construct suitable wave packets and analyze their dynamics from a quasi-classical region to the initial singularity. The appearance of a non singular dynamics comes out as far as the behavior of the probability density is investigated. Furthermore, reliable indications arise about the absence of a Big-Bounce, as predicted in recent issues of Loop Quantum Cosmology.
\end{abstract}


\maketitle 

\section{I. Introduction}

The classical theory of gravity, Einstein's General Relativity (GR), is very well tested on macroscopic scale, i.e. it is expected to work up to nuclear density. On the other hand this theory implies the very well-known singularity theorems \cite{HE}. The prediction of singularities undoubtedly represents a breakdown of GR in that its classical description cannot be expected to remain valid at the extreme condition near a spacetime singularity. One of the most important example is the Big-Bang singularity appearing in cosmological models. 

In this work we analyze the quantum features of the cosmological singularity in the framework of minimal length uncertainty relations. The existence of a fundamental minimal scale is long expected in a quantum theory of gravity and it invites the possibility that there are corrections to the usual Heisenberg uncertainty principle such it becomes what is known as a Generalized Uncertainty Principle (GUP). Interest in minimal length or Generalized Uncertainty Principle has been motivated by study on perturbative Sting Theory \cite{String}, considerations regarding the proprieties of black holes \cite{Mag} and de Sitter space \cite{Sny}. However, in recent years, a big amount of work has been made in this active field in a wide variety of directions (see for example \cite{GUP1} and the references therein) and here we present the first application to a minisuperspace dynamics.

In view applying the idea a cutoff scale to the minisuperspace, the GUP approach appears the best physically grounded. In fact, it relies on a modification of the canonical prescriptions for quantization and, in this respect, it can be reliable applied to any dynamical system. Thus, GUP formalism allow us to implement some peculiar features of String Theory in quantizing a cosmological model. The application to the cosmological minisuperspace of other noncommutative approaches (like the k-Minkowski formulation \cite{k-Min}) would require to address its Minkoskian nature as a physical one. But no reliable evidence arises that this Lorentian signature has a deep physical interpretations.        

The studied model is the isotropic flat Friedmann-Robertson-Walker (FRW) ($k=0$) cosmological model with a massless scalar field. By the minisuperspace reduction we are able to treat the scale factor as a coordinate of the particle (Universe), i.e. in the framework of GUP a nonzero minimal uncertainty in the isotropic volume of the Universe appears. Despite the simplicity of the model, this deserves interest for mainly two reasons: i) the scalar field can be regarded as an ``emergent time'' for the quantum evolution, ii) in the Wheeler-DeWitt framework the unavoidable classical singularity cannot be solved and the wave packet follow a classical trajectory up to the ``initial'' singularity \cite{Ish,APS}. In the GUP approach, as we will see, the situation is very different. In fact the wave packet, peaked at late times (at energies much smaller than the Planck's one), ``escape'' from the classical trajectory in the dynamics toward the cosmological singularity. Therefore the probability density to find the Universe near the classic time where the singularity appears goes to zero and, in some sense, our quantum Universe approach stationary states ``near the Planckian region''. In this sense the cosmological singularity is solved by the modified Heisenberg algebra.

The structure of the letter is the following. In Section II and III the quantum-mechanical structure which underlines the GUP framework and the canonical quantization of the FRW ($k=0$) cosmological model with a massless scalar field are briefly reviewed, respectively. Section IV is devoted to a discussion on the scalar field taken as a relational time. In Section V the ``generalized quantization'' of the model is performed and either the probability density to find the Universe near the classical singularity, either the wave packets dynamics are investigated in detail. In Section VI concluding remarks follow. 

\section{II. Quantum Mechanics with GUP}

In this section we briefly review some aspects and results of a nonrelativistic quantum mechanics with nonzero minimal uncertainties in position \cite{Kem}. We consider the Heisenberg algebra generated by $\bf q$ and $\bf p$ obeying the commutation relation (we work with $\hbar=c=1$ units) 
\be
[{\bf q},{\bf p}]=i(1+\beta{\bf {p}}^2), 
\ee
where $\beta$ is a ``cutoff'' constant. This commutation relation leads to the uncertainty relation
\be
\Delta q \Delta p\geq \f 1 2\left(1+\beta (\Delta p)^2+\beta \langle{\bf p}\rangle^2\right),
\ee
which appears in perturbative String Theory \cite{String}. This Generalized Uncertainty Principle implies an absolutely minimal uncertainty in position (considering $\langle{\bf p}\rangle=0$) $\Delta q_0=\sqrt\beta$. The existence of a nonzero uncertainty in position implies that there cannot by any physical state which is a position eigenstate since an eigenstate would of course have zero uncertainty in position. This consideration force us to work on the momentum space where we have the following representation:
\be\label{rep}
{\bf p}\psi(p)=p\psi(p), \qquad {\bf q}\psi(p)=i(1+\beta p^2)\p_p\psi(p).
\ee
In order to recover information on positions we have to study the states which realize the maximally allowed localization. Such states $\vert\psi^{ml}_{\zeta}\rangle$ of maximal localization around a position $\zeta$ have the proprieties $\langle\psi^{ml}_{\zeta}\vert {\bf q}\vert\psi^{ml}_{\zeta}\rangle=\zeta$ and $(\Delta q)_{\vert\psi^{ml}_{\zeta}\rangle}=\Delta q_0$ and obey the following equation
\be
\left({\bf q}-\langle{\bf q}\rangle + \f{\langle [{\bf x},{\bf p}]\rangle}{2(\Delta p)^2}\ {\bf p}\right){\vert\psi\rangle}=0,
\ee
where we remember that the absolutely maximal localization can only be obtained for $\langle{\bf p}\rangle=0$. Thus, the maximally localized states in the momentum space read explicitly as
\be
\psi^{ml}_{\zeta}(p)\sim\f 1 {(1+ \beta p^2)^{1/2}} \exp\left(-i\f{\zeta \tan^{-1}(\sqrt{\beta}p)}{\sqrt{\beta}}\right),
\ee
which, in the limit $\beta\rightarrow0$, reduce to the ordinary plane waves. At this point we can project an arbitrary state $\vert\psi\rangle$ on the maximally localized states $\vert\psi^{ml}_{\zeta}\rangle$, in order to obtain the probability amplitude for a particle being maximally localized around the position $\zeta$ (i.e. with standard deviation $\Delta q_0$). We call these projections the ``quasiposition wave function'' $\psi(\zeta)\equiv\langle\psi^{ml}_{\zeta}\vert\psi\rangle$; explicitly we have
\be\label{qwf} 
\psi(\zeta)\sim\int^{+\infty}_{-\infty}\f{dp}{(1+\beta p^2)^{3/2}} \exp\left(i\f{\zeta \tan^{-1}(\sqrt{\beta}p)}{\sqrt{\beta}}\right)\psi(p).
\ee
This is nothing but a generalized Fourier transformation, where in the limit $\beta\rightarrow0$ the ordinary position
wave function $\psi(\zeta) = \langle\zeta\vert\psi\rangle$ is recovered.  

\section{III. Canonical Quantization of the FRW model}

The canonical quantization of the homogeneous, isotropic, flat ($k=0$) cosmological model with a massless scalar field is reviewed (for more details see \cite{Ish,APS}). The Hamiltonian constraint for this model has the form
\be\label{con}
H_{grav}+H_{\phi}\equiv-9\kappa p_x^2x+\f3 {8\pi}\f{p_{\phi}^2}{x}\approx0 \quad x\equiv a^3,
\ee      
where $\kappa=8\pi G\equiv8\pi l_P^2$ is the Einstein constant and $a$ is the scale factor. In the classical theory, the phase space is $4$-dimensional, with coordinates $(x,p_x;\phi,p_{\phi})$. At $x=0$ the physical volume of the Universe goes to zero and the singularity appears. Since $\phi$ does not enter the expression of the constraint, $p_{\phi}$ is a constant of motion and therefore each classical trajectory can be specified in the $(x,\phi)$-plane. Thus $\phi$ can be considered as an internal time (see below) and the dynamical trajectory reads as
\be\label{clastra}
\phi=\pm\f 1 {\sqrt{24\pi\kappa}}\ln\left|\f x {x_0}\right|+\phi_0,
\ee
where $x_0$ and $\phi_0$ are integration constants. In this equation, the plus sign describes an expanding Universe from the Big-Bang, while the minus sign a contracting one into the Big-Crunch. We now stress that the classical cosmological singularity is reached at $\phi=\pm\infty$ and every classical solution, in this model, reaches the singularity. 

At quantum level the Wheeler-DeWitt equation, associated to the constraint (\ref{con}), tells us how the wave function $\Psi(x,\phi)$ evolves as $\phi$ changes; in this respect we can regard the argument $\phi$ of $\Psi(x,\phi)$ as an ``emergent time'' and the scale factor as the real physical variable. In order to have an explicit Hilbert space, we perform the natural decomposition of the solution  into positive and negative frequency parts. Therefore, the solution of this Wheeler-DeWitt equation has the very well-known form
\be\label{solcan} 
\Psi_\epsilon(x,\phi)=x^{-1/2}\left(Ax^{-i\gamma}+Bx^{i\gamma}\right)e^{i \sqrt{24\pi\kappa}\epsilon\phi}, 
\ee
where $\gamma=\f 12(4\epsilon^2-1)^{1/2}\geq0$ and $\epsilon^2$ being the eigenvalue of the operator $\Xi/24\pi\kappa$ defined below. Thus the spectrum is purely continuous and covers the interval $(\sqrt{3}/2l_P,\infty)$ \cite{Ish}. The wave function $\Psi_\epsilon(x,\phi)$ is of positive frequency with respect to the internal time $\phi$ and satisfies the positive frequency (square root) of the quantum constraint (\ref{con}); we deal with a Sch\"odinger-like equation $-i\p_\phi\Psi=\sqrt{\Xi}\Psi$, where $\Xi\equiv24\pi\kappa\hat{x}\hat{p_x}^2\hat{x}$. 

In order to examine the behavior of the classical singularity at quantum level we have to clarify a general criteria for determining whether the quantized models actually collapse \cite{Got}. Unfortunately there is not such a rigorous criteria yet. An early idea was to impose the condition that the wave function vanishes at the singularity $a=0$ \cite{DW}, but this boundary conditions has little to do with the quantum singularity avoidance. It seems better to study the expectation values of observables which classically vanish at the singularity. In fact, $\vert\Psi(a=0,t)\vert^2$ is merely a probability density and thus, for example, one might have an evolving state that ``bounces'' (i.e. a nonsingular wave packet), even if $\vert\Psi(a=0,t)\vert\neq0$ for all $t$. On the other hand, if one could find a wave packet so that the probability $P_{\delta}\equiv\int_0^\delta\vert\Psi(a,t)\vert^2da\simeq0$ for $\delta$ being a very small quantity, then one could reasonably claim to have a nocollapse situation.

Let's now come back to the canonical FRW model. It is not difficult to see that, in this framework, the unavoidable classical singularity is not tamed by quantum effects. In fact, if one starts with a state localized at some initial time, then its peak moves along the classical trajectory and falls into the classical singularity. Additionally, from the eigenfunctions (\ref{solcan}) it is clear that the probability defined above diverges indicating that the Wheeler-DeWitt formalism does not solve the classical singularity.

\section{IV. On the scalar field as a relational time}

Let us now discuss in some details the role of a matter field, in particular of a scalar field $\phi$, as definition of time for the quantum dynamics of the gravitational field. As well-known, in any diffeomorphism-invariant quantum field theory (like Quantum General Relativity), the Sch\"odinger equation is replaced by a Wheeler-DeWitt equation, in which the time coordinate disappears from the formalism. This is the so-called {\it problem of time} and it is one of the major conceptual puzzle in Quantum Gravity (for an exhaustive discussion on it see \cite{PT}). To overcome this difficulty, it was argued that the ``physical time'' has to be described using some of the canonical variables of the gravitational field. In particular it is possible to distinguish between an {\it intrinsic time}, which is constructed entirely from the 3-metric, and an {\it extrinsic time}, which is constructed from both the metric and the extrinsic curvature. However, it was shown how it is difficult to provide a satisfactory definition of time using both these concepts. Moreover, a physical clock has to be made of matter with certain definite proprieties. Therefore it seems natural to use a matter field to define a relational time. The concept of a relational time is principally based on the idea about the absence of the time at a fundamental level. Thus, the dynamics of a system can be described only on a relational point of view, i.e. a field has to be evolved with respect to another one (for more details on this approach see \cite{Ro04}). From this perspective, we will evolve the physical degrees of freedom of the gravitational field toward a massless scalar field and in this respect it will be treated as an
 external time. 

In particular, in quantum cosmology, the choice of a scalar field as an external time appears as the most natural one. In fact, a monotonic behavior near the classical singularity, of $\phi$ as function of the scale factor (more precisely the variable which describe the isotropic expansion of the Universe) always appears. Let us consider the case of the Bianchi IX model, i.e. the most general (together with Bianchi VIII) homogeneous cosmological model, in the presence of a massless scalar field $\phi$. In such a case, the Hamiltonian constraint $H_{grav}+H_{\phi}\approx0$, has the form
\be
\kappa\left[-\f{p_a^2}a+\f1{a^3}(p_+^2+p_-^2)\right]-\f a{4\kappa}V(\beta_\pm)+\f3 {8\pi}\f{p_{\phi}^2}{a^3}\approx0,
\ee
where the $a$ variable describes the isotropic Universe expansion, $\beta_\pm$ are associated to the space anisotropies and the potential term $V(\beta_\pm)$ accounts for the spatial curvature of the model. Since the determinant of the 3-metric is given by $h=a^3$, it is easy to recognize that the classical singularity appears for $a\rightarrow0$. Therefore near the cosmological singularity, i.e. in a purely quantum region, the potential term can be neglected (for a discussion on the consistency condition which ensures that the quasi-classical limit of the Universe dynamics is reached before the potential term becomes important see \cite{BM06}). As before, we will regard the scalar field as the time for the dynamics and, in this respect, the equations of motion of the system read as 
\be\label{monrel}
a(\phi)=B\exp\left(\f{A\phi}{\sqrt{A^2-c}}\right),\, p_a(\phi)=\f AB\exp\left(-\f{A\phi}{\sqrt{A^2-c}}\right),
\ee
where $A$ and $B$ are integration constants and $c\equiv p_+^2+p_-^2=const$ ($p_+=const$ and $p_-=const$). 

An important point has now to be stressed. As well-known \cite{BKL82}, the dynamics toward the cosmological singularity of a generic inhomogeneous Universe is described point by point, by that one of a Bianchi IX model. More precisely, it is possible to show that the spatial point, toward the singularity, dynamically decouple and therefore the spatial geometry can be viewed as a collection of small patches, each of which evolves independently as a Bianchi IX model. In this respect, the above monotonic relation (\ref{monrel}) between a massless scalar field and the isotropic variable of the Universe, is a real general feature of the gravitational field and this clarify the choice of a scalar field as a relational time. Therefore we can conclude that, not only in the symmetric model treated in this work but also in the most general cosmological model, the scalar field can be regarded as a good time for the gravitational dynamics. 

\section{V. Generalized Quantization of the FRW model}

We will analyze the quantization of the FRW ($k=0$) model in the framework of minimal length uncertainty relation. As in the canonical case, let us decompose the solution of the ``generalized Wheeler-DeWitt equation'' into positive and negative frequency parts and focus on the positive frequency sector. Remembering that we have to work in the momentum space, the wave function reads as: $\overline{\Psi}(p,\phi)=\Psi(p)e^{i \sqrt{24\pi\kappa}\epsilon\phi}$, where from now on $p\equiv p_x$. Thus, as soon as we regard the scalar field as an ``emergent time'' for the quantum evolution, then we treat in the ``generalized'' way only the real degree of freedom of the problem: the isotropic volume $x$. Therefore the quantum equation relative to the Hamiltonian constraint (\ref{con}), considering the representation (\ref{rep}), is the following
\be\label{emu}
\mu^2(1+\mu^2)^2\f{d^2\Psi}{d\mu^2}+2\mu(1+\mu^2)(1+2\mu^2)\f{d\Psi}{d\mu}+\epsilon^2\Psi=0,
\ee  
where we have defined a dimensionless parameter $\mu\equiv\sqrt\beta p$. In order to integrate the above equation we introduce the variable $\rho\equiv\tan^{-1}\mu$, which maps the region $0<\mu<\infty$ to $0<\rho<\pi/2$. Then, performing another change of variables: $\xi\equiv\ln(\sin\rho)$ ($-\infty<\xi<0$), equation (\ref{emu}) reduces to
\be
\f{d^2\Psi}{d\xi^2}+2\left(\f{1+e^{2\xi}}{1-e^{2\xi}}\right)\f{d\Psi}{d\xi}+\epsilon^2\Psi=0,
\ee 
which can be explicitly solved and whose general solution reads as
\be
\Psi_\epsilon(\xi)=C_1 e^{-\xi(1+\alpha)}\left(1+b_+e^{2\xi}\right)+C_2 e^{-\xi(1-\alpha)}\left(1+b_-e^{2\xi}\right),
\ee
where $\alpha=\sqrt{1-\epsilon^2}$ and $b_\pm=1\pm\alpha/(1\mp\alpha)$. At this point we have to analyze the ``quasiposition wave function'' relative to this problem in order to make a first comparison with the canonical case, in particular with the wave function (\ref{solcan}). In agreement with formula (\ref{qwf}) we have
\begin{multline}\label{quapos}
\Psi_\epsilon(\zeta)=\int_{-\infty}^0d\xi \exp{\left(\xi+i\zeta\tan^{-1}\left(\f{e^\xi}{\sqrt{1-e^{2\xi}}}\right)\right)}\\
\times\left[C_1 e^{-\xi(1+\alpha)}\left(1+b_+e^{2\xi}\right)+C_2 e^{-\xi(1-\alpha)}\left(1+b_-e^{2\xi}\right)\right],
\end{multline}
where $\zeta$, in this case, is expressed in units of $\sqrt\beta$. Thus we can easily see that our ``quasiposition wave function'', i.e. the probability amplitude for the particle (Universe) being maximally localized around the position $\zeta$, is nondiverging for all $\zeta$, as soon as we take the condition $C_1=0$. We stress that the canonical wave function (the function (\ref{solcan})) is diverging at the classical singularity $x=0$.

To get a better feeling with our quantum Universe we construct and examine the motion of wave packets. Let's now construct states peaked at late times
\be\label{wp} 
\Psi(\zeta,t)=\int_0^\infty d\epsilon g(\epsilon)\Psi_\epsilon(\zeta)e^{i\epsilon t},
\ee
where we have defined the dimensionless time $t=\sqrt{24\pi\kappa}\phi$. In the following we take $g(\epsilon)$ to be a Gaussian distribution peaked at some $\epsilon^\ast\ll1$, which corresponds to be peaked at energy much less then the Plank energy $1/l_P$ (we recall that $\epsilon\sim\mathcal{O}(\overline\epsilon\ l_P)$, where $\overline\epsilon$ have dimension $1$ in energy). The analytic computation of the integral (\ref{wp}) for the wave function (\ref{quapos}) is impossible to perform. So, in order to describe the motion of wave packets we have to evaluate (\ref{wp}) numerically.

At first, we want to analyze the most interesting region, i.e. where $\zeta\simeq0$, which corresponds to the purely quantum region, where the physical volume is Planckian. In fact, if we put $\beta\sim\mathcal{O}(l_P^6)$, the minimal uncertainty in position is of order of the Planckian volume. The ``quasiposition wave function'' (\ref{quapos}) can be expanded in order to give the probability density around $\zeta\simeq0$: $\vert\Psi(\zeta,t)\vert^2\simeq\vert A(t)\vert^2+\zeta^2\vert B(t)\vert^2$. Therefore, starting with a state peaked at some $\epsilon^\ast\ll1$, the probability density of finding the Universe ``around the Planckian region'' is $\vert A(t)\vert^2$, where $A(t)$ reads
\be\label{At}
A(t)=2C_2\int_0^\infty d\epsilon\f{(1+2\sqrt{1-\epsilon^2})e^{-\f{(\epsilon-\epsilon^\ast)^2}{2\sigma^2}+i\epsilon t}}{\sqrt{1-\epsilon^2}\left(3-\epsilon^2+3\sqrt{1-\epsilon^2}\right)} .
\ee
We evaluate the above integral numerically for $\epsilon^\ast=10^{-3}$, $\sigma^2=1/20$ and we take the constant $2C_2=1$. The probability density $\vert A(t)\vert^2$ is very well approximated by a Lorentzian function (see Fig. 1). As we can see form the picture, this curve is peaked around $t=0$. This value corresponds to the classical time for which $x(t)=x_0$ (in (\ref{At}) we consider $t_0=0$). Thus, for $x_0\sim\mathcal{O}(l_P^3)$, the probability density to find the Universe in a Planckian volume is peaked around the corresponding classical time. As a matter of fact this probability density vanishes for $t\rightarrow-\infty$, where the classical singularity appears. This is the meaning when we claim that the classical cosmological singularity is solved by this model. 

In order to describe the motion of the wave packet we evaluate $\vert\Psi(\zeta,t)\vert^2$ from the integral (\ref{wp}) of the wave function (\ref{quapos}). As before, we consider a wave packet initially peaked at late times and let it evolve numerically ``backward in time''. We use the same parameters for the integration performed above. The result of the integration is that the probability density, at different fixed values of $\zeta$, is very well approximated by a Lorentzian function yet. The width of this function remains, actually, the same as the states evolves from large $\zeta$ ($10^3$) to $\zeta=0$. For all fixed $t$ the probability density is well-fixed by a Lorentzian function and the width of these function, also in this case, remain almost the same during the evolution. These states are sharply peaked for $\zeta\sim\mathcal{O}(1)$ (which in our units correspond to $\zeta\sim\mathcal{O}(\sqrt\beta)\sim\mathcal{O}(l_P^3)$). The peaks of Lorentzian functions, at different $\zeta$ values, move along the classically expanding trajectory (\ref{clastra}) for values of $\zeta$ larger then $\sim4$. Near the Planckian region, i.e. when $\zeta\in[0,4]$, we observe a modification of the trajectory of the peaks. In fact they follow a power-law up to $\zeta=0$, reached in a finite time interval and ``escape'' from the classical trajectory toward the classical singularity (see Fig. 2). The peaks of the Lorentzian at fixed time $t$, evolves very slowly remaining close to the Planckian region. Such behavior outlines that the Universe has a stationary approach to the cutoff volume, accordingly to the behavior in Fig. 2.  

This peculiar behavior of our quantum Universe is different from other approaches to the same problem. In fact, recently, it was shown how the classical Big-Bang is replaced by a Big-Bounce in the framework of Loop Quantum Cosmology (LQC) \cite{APS}. Intuitively, one can expect that the bounce and so the consequently repulsive features of the gravitational field in the Planck regime are consequences of a Planckian cut-off length. But this is not the case. As matter of fact we can observe from Fig. 2 that there is not a bounce for our quantum Universe. The main differences between the two approaches resides in the quantum modification of the classical trajectory. In fact, in the LQC framework we observe a ``quantum bridge'' between the expanding and contracting Universes; in our approach, contrarily, the probability density of finding the Universe reaches the Planckian region in a stationary way.
\begin{figure}
\begin{center}
\includegraphics[height=2in]{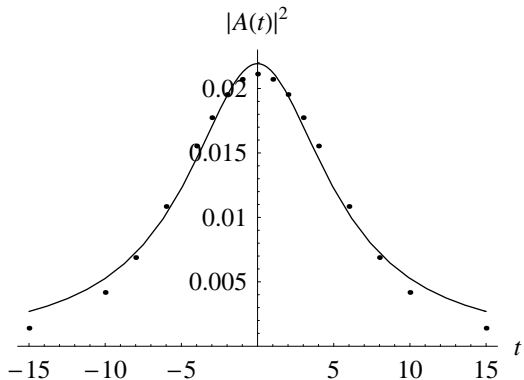}
\caption{The points represent the result of the numerical integration and are fitted by a Lorentzian $L(t)=0.692/(t^2+31.564)$ heaving width, at the inflection point, $3.243$.} %\label{Fig2}
\end{center}
\end{figure}
\begin{figure}
\begin{center}
\includegraphics[height=2in]{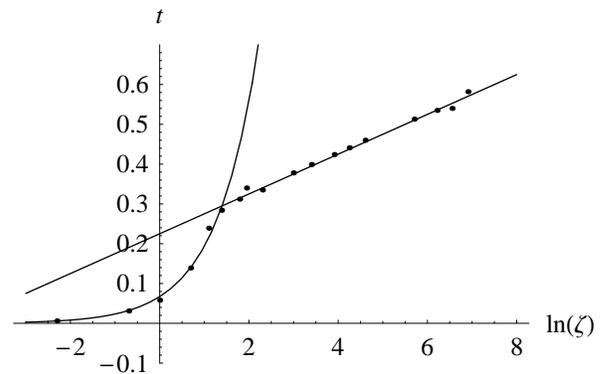}
\caption{The peaks of the probability density $\vert\Psi(\zeta,t)\vert^2$ are plotted as functions of $t$ and $\ln(\zeta)$. The points (resulting from numerical computation) are fitted by a logarithm $0.050\ln(\zeta)+0.225$ for $\zeta\geq4$ and by a power law $0.067\zeta^{1.060}$ for $\zeta\in[0,4]$.} %\label{Fig2}
\end{center}
\end{figure}  

\section{VI. Concluding Remarks}

In this work we have shown the effects of a modified Heisenberg algebra, which reproduces a GUP as appeared in studies on String Theory \cite{String}, on the cosmological singularity. Two main remarks on our analysis must be stressed: i) we have restricted our investigation to an isotropic, homogeneous model, although a quantum Universe has to be described by a generic cosmological model where all the symmetries are removed; ii) the width of the Lorentzian functions in $t$, and so the probability density, are too wide to formulate a precise description of the wave packets toward the Planck era. These shortcomings of our analysis will be addressed in a more detailed discussion on this topics, which will appear elsewhere.

In this model, the resolution of the classical singularity relies on the mathematical framework used. In fact, the features of the ``quasiposition wave function'' are responsible for the possibility to expand the probability density $\vert\Psi(\zeta,t)\vert^2$ near $\zeta\simeq0$, while the corresponding probability density near $a\simeq0$ diverges in the Wheeler-DeWitt case.  

Certainly, the main result of the present work is that in the framework of GUP no evidence for a Big-Bounce seems to come out. In fact the behavior of the wave packet near the Planckian region does not allow for a bounce, in the sense that the peak of the probability density approaches the region $\zeta\simeq0$ by a power-law and it does not rejoin itself to the contracting region. The big width of the probability density is not qualitatively a real problem for the singularity absence. In fact, near $\zeta\simeq0$, the behavior we have found is not the classical logarithmic one even taking the error into account.    

\section{Acknowledgments}
We would like to thank Roberto Guida for his help in generating the manuscript plots.

\end{document}